\documentclass[reprint,longbibliography]{revtex4-1}

\usepackage{color}
\usepackage{graphicx}
\usepackage{mathrsfs}
\usepackage{paralist}
\usepackage{enumerate}
\usepackage{amssymb,amsfonts,amsmath}

\newcommand{\blue}[1]{{\color{black} #1}}
\begin{document}


\title{Fluid-solid-electric lock-in of energy-harvesting piezoelectric flags} 



\author{Yifan \surname{Xia}}
\email[]{xia.yifan@ladhyx.polytechnique.fr}
\affiliation{LadHyX--D\'epartement de M\'ecanique, \'Ecole Polytechnique, Route de Saclay, 91128 Palaiseau, France}
\author{S\'ebastien \surname{Michelin}}
\affiliation{LadHyX--D\'epartement de M\'ecanique, \'Ecole Polytechnique, Route de Saclay, 91128 Palaiseau, France}
\author{Olivier \surname{Doar\'e}}
\affiliation{ENSTA Paristech, Unit\'e de M\'ecanique (UME), 828 boulevard des Mar\'echaux, 91762, Palaiseau, France}


\date{\today}

\begin{abstract}
The spontaneous flapping of a flag in a steady flow can be used to power an output circuit using piezoelectric elements positioned at its surface. Here, we study numerically the effect of inductive circuits on the dynamics of this fluid-solid-electric system and on its energy harvesting efficiency. In particular, a destabilization of the system is identified leading to energy harvesting at lower flow velocities. Also, a frequency lock-in between the flag and the circuit is shown to significantly enhance the system's harvesting efficiency. These results suggest promising efficiency enhancements of such flow energy harvesters through the output circuit optimization.
\end{abstract}

\keywords{energy harvesting; piezoelectric material; fluid-structure instability; resonant circuit; lock-in }
\maketitle 

\section{Introduction}
Flow induced instabilities and vibrations have recently received a renewed attention as potential mechanisms to produce electrical energy from geophysical flows (wind, tidal currents, river flows, etc.). They indeed enable a spontaneous self-sustained motion of a solid body which can be used as a generator, effectively converting this mechanical energy into electrical form \cite{bernitsas:2008,frayne:2009,frayne:2011, xiao:2014,grouthier:2014}.

A canonical example of such instability is the flapping of a flexible plate in an axial flow (e.g. a flag), which has been extensively investigated for its rich and complex dynamics \cite{shelley:2011}. The origin of this instability lies in a competition between the destabilizing fluid force and the stabilizing structural stiffness. Beyond a critical flow velocity $U_c$, the flag becomes unstable, leading to large amplitude self-sustained flapping \cite{watanabe:2002a, argentina:2005, eloy:2007, connell:2007, eloy:2008, alben:2008c, michelin:2008, virot:2013}.

Energy harvesting based on flapping plates may follow two routes: producing energy either from the displacement \cite{tang:2008b} or from the deformation of the plate \cite{taylor:2001, allen:2001, singh:2012}. The latter has recently been the focus of several studies based on active materials \cite{giacomello:2011}. 

Piezoelectric materials, considered in this \blue{article}, produce electric charge displacement when strained \cite{yang:2005}, a ``direct piezoelectric effect'' that effectively qualifies them as electric generators. This electric charge can be used in an output circuit connected to their electrodes, as in vibration control applications \cite{hagood:1991}. Piezoelectric materials also introduce a feedback coupling of the circuit onto the mechanical system: any voltage between the electrodes creates an additional structural stress that modifies its dynamics (inverse piezoelectric effect).

\blue{
The concept of piezoelectric energy generator has received an increasing amount of interest in the last 20 years \cite{williams:1996, umeda:1996}. Its basic idea is to convert ambient vibration energy to useful electric energy through piezoelectric materials implemented on vibration sources. Many researchers have contributed to this field in order to improve the efficiency of such energy-harvesting systems \cite{sodano:2004, anton:2007, erturkbook:2011, calio:2014piezoelectric}. Some studies show that simple resonant circuits, {i.e.} resistive-inductive circuits combined with the piezoelectric material's intrinsic capacitance \cite{yang:2005}, offer promising opportunities to achieve high efficiency \cite{shenck:2001, demarquis:2011}.

Flow energy harvesting can be achieved by exploiting the unsteady forcing of the vortex wake generated by an upstream bluff body to force the deformation of a piezoelectric membrane \cite{allen:2001, taylor:2001, podering:2004}. Fluid-solid instabilities offer a promising alternative as they are able to generate  spontaneous and self-sustained structural deformation of the piezoelectric structure, {e.g.} cross-flow instabilities \cite{kwon:2010, demarquis:2011, dias:2013}. In their work, De Marquis {\it et al.} \cite{demarquis:2011} used a resistive circuit and a resistive-inductive one, and found in addition to the beneficial effect of the resonance to the energy harvesting, that a resistive-inductive circuit may also affect the stability of the vibration source. However, the resonant circuit's influence on the structure's dynamics was not reported in this work.

Fluid-solid instabilities in axial flows, including the aforementioned flapping flag instability, are also studied in the  context of piezoelectric energy-harvesting \cite{dunnmon:2011, doare:2011, akcabay:2012, michelin:2013}. In particular, Michelin \& Doar\'e \cite{doare:2011, michelin:2013} considered a piezoelectric flag coupled with a purely resistive output. They observed moderate efficiency, which is maximized when the characteristic timescale of the circuit is tuned to the frequency of the flag. A significant impact of the circuit's properties on the fluid-solid dynamics was also identified.
}

The present work therefore focuses on the coupling of the fluid-solid system, \blue{i.e. the flapping piezoelectric flag, to a basic resonant circuit (resistive-inductive loop with the piezoelectric material's intrinsic capacitance).} Resonance is expected when the flapping frequency $\omega$ of the flag, forcing the circuit, matches \blue{the circuit's} natural frequency. Using linear stability analysis and nonlinear numerical simulations of the fluid-solid-electric coupled system, we investigate the impact of such resonance on the dynamics and on the amount of energy that can be extracted from the device (i.e. the energy dissipated in the resistive elements). This explicit description of both the fluid-solid and the electric systems' dynamics provides a deeper and more accurate insight into the energy harvesting process than its classical modeling as a pure damping \cite{tang:2008b, xiao:2014, singh:2012}.

\section{Fluid-solid-electric model}
\begin{figure}
  \centering
  \includegraphics{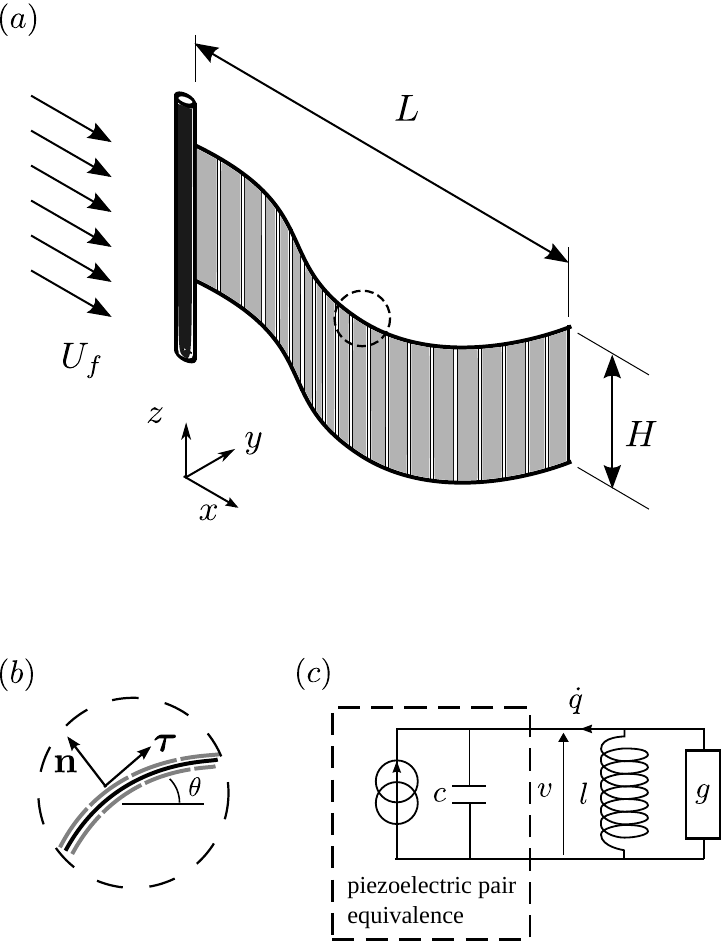}
  \caption{\label{fig:flags_piezo_circuit}Flapping piezoelectric flag in a uniform axial flow. $(a)$: 3-dimension view; $(b)$: zoom of the circled area in $(x,y)$ plane; $(c)$: equivalent circuit of a piezoelectric pair connected with a parallel RL circuit.} 
\end{figure}

\blue{
\subsection{Fluid-solid coupling}
}
The coupled system considered here is a cantilevered plate of length $L$ and span $H$, placed in an axial flow of density {$\rho_f$} and velocity $U_f$. The flag's surface is covered by pairs of piezoelectric patches (Fig.~\ref{fig:flags_piezo_circuit}$a$). Within each pair, two patches of reversed polarities are connected through the flag, {the} remaining electrodes being connected to the output circuit \cite{bisegna:2006, doare:2011}. The resulting three-layer sandwich plate is of lineic mass $\mu$ and bending rigidity $B$. We restrict here to purely planar deformations (bending in the $z$-direction and twisting are neglected). The flag's dynamics {are} described using an inextensible Euler--Bernoulli beam model forced by the fluid:
\blue{
\begin{align}
\label{EB_dimensional}
\mu\mathbf{\ddot X}&=(\mathcal{T}\boldsymbol{\tau}-\mathcal{M}'\mathbf{n})'
+\mathbf{F}_\text{f},\\
\label{Inextens}
\mathbf{X}'&=\boldsymbol{\tau}
\end{align}
with clamped-free boundary conditions:
\begin{align}
&\text{at}\quad s=0:~\mathbf{X}=\mathbf{\dot X}=0\label{BCd_1},\\
&\text{at}\quad s=L:~\mathcal{T}=\mathcal{M}=\mathcal{M}'=0.\label{BC_2}
\end{align}
Here $\mathcal{T}$ and $\mathcal{M}$ are respectively the tension and the bending moment. Throughout this article, $~\dot{ }~$ and $~{ }'~$ denote derivatives with respect to $t$ and $s$, respectively. The fluid loading $\mathbf{F}_\text{f}$ is computed using a local force model from the relative velocity of the flag to the incoming flow:
\begin{equation}
U_n\mathbf{n}+U_\tau\boldsymbol{\tau}=\dot{\mathbf{X}}-U_f\mathbf{e}_x.\label{u_rela}
\end{equation}
The present fluid model includes two different contributions. The first one results from the advection of the fluid added momentum by the flow, an inviscid effect, and can be obtained analytically in the slender body limit through the \blue{{\it Large Amplitude Elongated Body Theory}} \cite{lighthill:1971}:
\blue{
\begin{equation}
\label{f_react}
\mathbf{F}_\text{react}=-m_a\rho H^2\left[ \dot U_n-(U_n U_\tau)'+\frac{1}{2}U_n^2\theta' \right]\mathbf{n}.
\end{equation}}
Candelier {\it et al} \cite{candelier:2011} recently proposed an analytic proof of this result, and successfully compared it to RANS simulations for fish locomotion problems. These authors also stated that in the case of spontaneous flapping, it is necessary to account for the effect of lateral flow separation, which is empirically modeled by the following term \cite{taylor:1952}:
\blue{\begin{equation}
\label{f_resis}
\mathbf{F}_\text{resist}=-\frac{1}{2}\rho HC_d|U_n|U_n\mathbf{n},
\end{equation}
where $C_d$ is the drag coefficient for a rectangular plate in transverse flow. It is important to mention that the flapping flag dynamics considered here implies a large Reynolds number. A 10 cm long/wide flag in a wind flowing at around $5$ m/s leads to $Re\sim 10^4$, while in water we would have $Re\sim 10^5$ for a flow velocity around $1$ m/s. These values of $Re$ are sufficiently large to justify a constant value of $C_d$ \cite{munson:2012fundamentals}. A $Re$-dependence could however be introduced to extend the applicability of this model to intermediate $Re$.

The fluid forcing is the sum of these two terms:
\begin{equation}
\mathbf{F}_\text{f}=\mathbf{F}_\text{react}+\mathbf{F}_\text{resist}.
\end{equation}}The applicability of this result to flapping flag was confirmed experimentally, at least up to an aspect ratio $H/L=0.5$, the value considered in our work \cite{eloy:2012}.}

\blue{
\subsection{Piezoelectric effects}
Piezoelectric patch pairs are positioned on either side of the plate, with opposite polarity. This guarantees that, during the plate's bending, the two patches reinforce each other, rather than cancel out.} Each piezoelectric pair is connected to a resistance and an inductance in parallel connection (Fig.~\ref{fig:flags_piezo_circuit}$c$).

We focus on the limit of continuous coverage by infinitesimal piezoelectric pairs \cite{doare:2011,michelin:2013}. Within this limit, the electric state of the piezoelectric pairs is characterized by the local voltage $v$ and lineic charge transfer $q$, which are continuous functions of the streamwise Lagrangian coordinate $s$. The electrical circuits are characterized by a lineic conductance $g$, and a lineic inductive admittance 1/$l$. The electrical charge displacement across a piezoelectric pair resulting from the direct piezoelectric effect is given by:
\begin{equation}
 q=\chi\theta'+cv\label{direct_effect},
\end{equation}
where $\chi$ is a mechanical/piezoelectric conversion coefficient and $c$ is the lineic intrinsic capacity of a piezoelectric pair \cite{thomas:2009}. Equation~\eqref{direct_effect} shows that the effect on the circuit of the piezoelectric components is that of a current generator with an internal capacitance (Fig.~\ref{fig:flags_piezo_circuit}$c$). The charge conservation of the resulting RLC circuit leads to:
\begin{equation}
 v+gl\dot{v}+l\ddot{q}=0\label{circuit_RLC}.
\end{equation}
The inverse piezoelectric effect manifests as an added bending moment, so that the total bending moment in the structure is \blue{given by \cite{bisegna:2006}:
\begin{equation}
\label{moment_piezo}
\mathcal{M}=B\theta'-\chi v.
\end{equation}

Finally we define the harvested energy as the time average of the total rate of dissipation in the \blue{resistive elements}, which is formally given as:
\begin{equation}
\label{Q}
\mathscr{P}=\left<\int_0^Lgv^2\mathrm{d}s\right>,
\end{equation}
and the efficiency is defined as:
\begin{equation}
\eta=\frac{\mathscr{P}}{\mathscr{P}_\text{ref}}.
 \label{eta_def}
\end{equation}
$\mathscr{P}_\text{ref}$ is the kinetic energy flux of fluid passing through the cross section occupied by the flag:
\begin{equation}
\label{p_ref}
\mathscr{P}_\text{ref}=\frac{1}{2}\rho U_\infty^3\times \mathscr{A}H,
\end{equation}
where $\mathscr{A}$ is the peak-to-peak amplitude of the flapping flag.
}

In the following, the problem is nondimensionalized using the elastic wave velocity $U_s=\sqrt{B/L^2\mu}$ as characteristic velocity. $L$, $L/U_s$, $\rho HL^2$, $U_s\sqrt{\mu/c}$, $U_s\sqrt{\mu c}$ are respectively used as characteristic length, time, mass, voltage and lineic charge. As a result, six non-dimensional parameters characterize the coupled system:
\begin{equation}
\begin{aligned}
M^*&=\frac{\rho_fHL}{\mu},&\quad U^*&=\frac{U_f}{U_s},&\quad H^*&=\frac{H}{L},\\
\alpha&=\frac{\chi}{\sqrt{Bc}},&\quad\beta&=\frac{c U_s}{gL},&\quad\omega_0&=\frac{L}{U_s\sqrt{lc}},\label{ParaAd}
\end{aligned}
\end{equation}
with $M^*$ the fluid-solid inertia ratio, $U^*$ the reduced flow velocity, and $H^*$ the aspect ratio. The piezoelectric coupling coefficient, $\alpha$, characterizes the fraction of the strain energy transferred to the circuit, and as such critically impacts the energy harvesting performance. Finally $\beta$ and $\omega_0$ characterize respectively the resistive and inductive properties of the circuit.

\blue{The effect of the mechanical parameters, $M^*$, $U^*$ and $H^*$, on the flapping flag dynamics have been extensively studied in the literature \cite{doare:2011, michelin:2013}.} In the following, we focus specifically on the dynamical properties of the circuit and maintain $M^*=1$ and $H^*=0.5$ throughout this study. \blue{Unless stated otherwise, we will also consider} $\alpha=0.3$, a value consistent with existing material properties \cite{doare:2011}. \blue{The effect of varying the piezoelectric coupling will also be briefly discussed.} The full nonlinear dynamics of the coupled system are now described in non-dimensional form by:
\blue{\begin{gather}
\mathbf{\ddot x}
=M^*(T\boldsymbol{\tau})'-\left(\theta''\mathbf{n}\right)'+\alpha\left( v'\mathbf{n} \right)'+M^*f_\text{f}\mathbf{n},\label{SysAd1}\\
\mathbf{x}'=\boldsymbol{\tau},\label{SysAd3}\\
\beta\ddot v+\dot v+\beta\omega_0^2v+ \alpha\beta\ddot\theta' =0,\label{SysAd2}
\end{gather}}and the \blue{nondimensional boundary conditions are:}
\blue{
\begin{align}
&\text{at}\quad s=0:~\mathbf{x}=\mathbf{\dot x}=0\label{BCad_1},\\
&\text{at}\quad s=1:~T=\theta'-\alpha v=\theta''-\alpha v'=0.\label{BCad_2}
\end{align}
}\blue{ The non-dimensional tension $T$ is computed using the inextensibility of the beam \cite{alben:2009a}}.

Finally the nondimensional fluid loading is obtained as:
\begin{equation}
f_\text{f}=-m_aH^*\left[ \dot u_n-(u_n u_\tau)'+\frac{1}{2}u_n^2\theta' \right]
-\frac{1}{2}C_d|u_n|u_n,\label{SysAdForce}
\end{equation}
with $m_a=\pi/4$ and $C_d=1.8$, the added mass and drag coefficients for a rectangular plate in transverse flow \cite{blevins:1990, buchak:2010}.

\section{Critical velocity}
The critical velocity $U^*_c$ is defined as the minimum flow velocity above which self-sustained flapping can develop and energy can be harvested. \blue{In this part, the influence of the RL loop on $U^*_c$ is studied using linear stability analysis, in the limit of small vertical displacement, {i.e.} $y\ll 1$, allowing linearization of Eqs.~ \eqref{SysAd1}--\eqref{SysAd2}. The resulting linear equations are:
 \begin{gather}
 \left(1+M_a\right)\ddot y+2M_aU^{*}\dot y'+M_aU^{*2}y''+y''''-\alpha v''=0\label{lin_beam},\\
  \beta\ddot v+\dot v+\beta\omega_0^2v+\alpha \beta\ddot y''=0,\label{lin_elec}
\end{gather}
where $M_a=\pi M^*H^*/4$.

Eqs.~\eqref{lin_beam} and \eqref{lin_elec} are then projected onto the fundamental beam modes and their second derivatives, respectively, and recast as an eigenvalue problem. The coupled system is unstable if one of its eigenfrequencies has a positive imaginary part.} 

\begin{figure}
\centering
\includegraphics{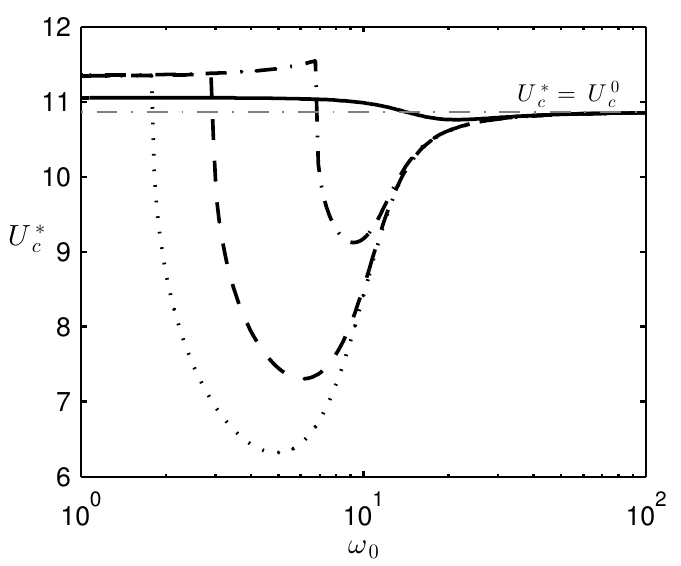}
\caption{\label{fig:thrs_M1_4beta}Evolution of the critical velocity with $\omega_0$ at $\alpha=0.3$ and $\beta=0.05$ (solid), $\beta=1$ (dash-dot), $\beta=4$ (dashed), $\beta=8$ (dotted). $U^*_c=U^0_c$ is plotted (dash-dot, gray) as a reference.}
\end{figure}

\blue{The evolution of $U^*_c$} with $\omega_0$ is computed using linear stability analysis and is shown on Fig.~\ref{fig:thrs_M1_4beta}. For intermediate values of $\omega_0$ ($3<\omega_0<10$), we observe a significant destabilizing effect of inductance that increases with $\beta$, as the circuit becomes dominated by inductive effects. For small $\beta$, however, no such destabilization is observed, as the inductance plays little role in this resistive limit. These results highlight a major benefit of the circuit's inductive behavior: the instability threshold may be lowered, resulting in energy harvesting with slower flow velocity. For $\omega_0 \gg 1$, the inductance acts as a short-circuit, and $U^*_c$ converges, regardless of $\beta$, to $U_c^0$, the critical flow velocity without 
coupling ($\alpha=0$). For $\omega_0\ll 1$, the 
effects of inductance are negligible, and $U_c^*\geqslant U_c^0$, illustrating the stabilizing effect of the resistance. Note that a destabilizing effect of the resistance can be observed at higher values of $M^*$\cite{doare:2011}, and in more general cases of damping \cite{doare:2010}.

\begin{figure}
\centering
\includegraphics{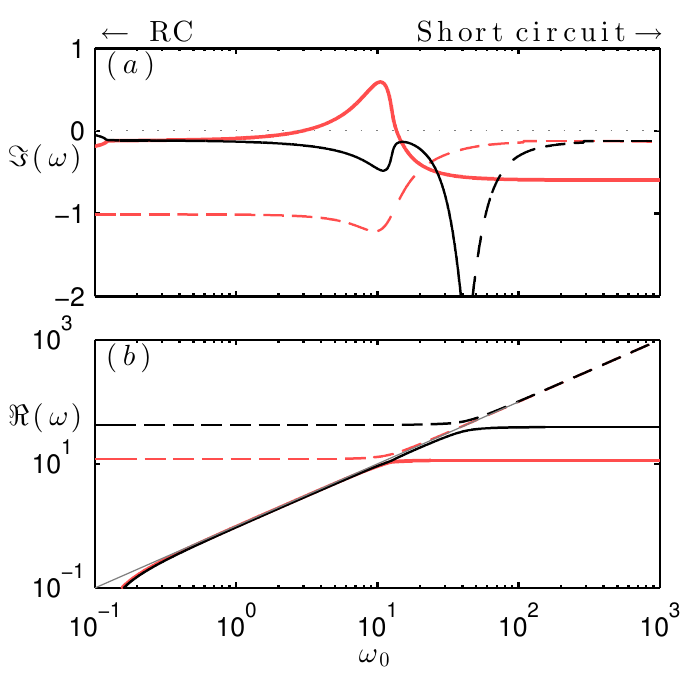}
\caption{(color online) Evolution with $\omega_0$ of $(a)$ the imaginary part (growth rate) and $(b)$ the real part (frequency) of the second and third pairs of eigenvalues for $\alpha=0.3$, $U^*=10$ and $\beta=4$. In both $(a)$ and $(b)$, solid lines represent mechanical modes and dashed lines represent electrical modes. The destabilized pair is plotted in red and the other plotted in black. In $(b)$, $\omega_0$ is plotted with thin gray solid line as a reference.}
\label{fig:EVs_destab}
\end{figure}

To determine the origin of this destabilization, Fig.~\ref{fig:EVs_destab} shows the evolution of the two most unstable pairs of eigenvalues with $\omega_0$ at $U^*=10$, which is lower than $U^0_c$ but higher than the minimum critical velocity (Fig.~\ref{fig:thrs_M1_4beta}). Starting from $\omega_0\gg 1$ and decreasing $\omega_0$, the electrical circuit evolves successively from short circuit, to RLC loop, and finally to a purely resistive circuit. Instability occurs when the imaginary part of any mode becomes positive.

In the absence of coupling ($\omega_0\gg 1$), both pairs consist of two eigenvalues: (i) an electrical mode with a frequency equal to $\omega_0$; (ii) a mechanical mode, with a frequency independent of $\omega_0$. Decreasing $\omega_0$ leads to interactions between the electrical and mechanical modes, successively within each pair. This interaction destabilizes the mechanical mode, leading to the flag's instability at intermediate $\omega_0$ (Fig.~\ref{fig:EVs_destab}). Note that this interaction within other pairs also leads to an increase of $\Im(\omega)$ for the mechanical mode, but does not lead to instability (at least for $M^*=1$). 

\section{Nonlinear dynamics and energy harvesting}
Above the critical velocity, the unstable coupled system experiences an exponential growth in its amplitude, which eventually saturates due to nonlinear effects. A direct integration of the fluid-solid-electric system's nonlinear equations \blue{Eqs.~\eqref{SysAd1}--\eqref{SysAd2}} is performed \blue{using an implicit second order time-stepping scheme \cite{alben:2009a}. The flag is meshed using Chebyshev-Lobatto nodes, and a Chebyshev collocation method is used to compute spatial derivatives and integrals. At each time step, the resulting nonlinear system is solved using Broyden's method \cite{broyden:1965}. The simulation is started with a perturbation in the flag's orientation ($\theta(s,t=0)\neq 0$), and is carried out over a sufficiently long time frame so as to reach a permanent regime.}

\begin{figure}
\centering
\includegraphics{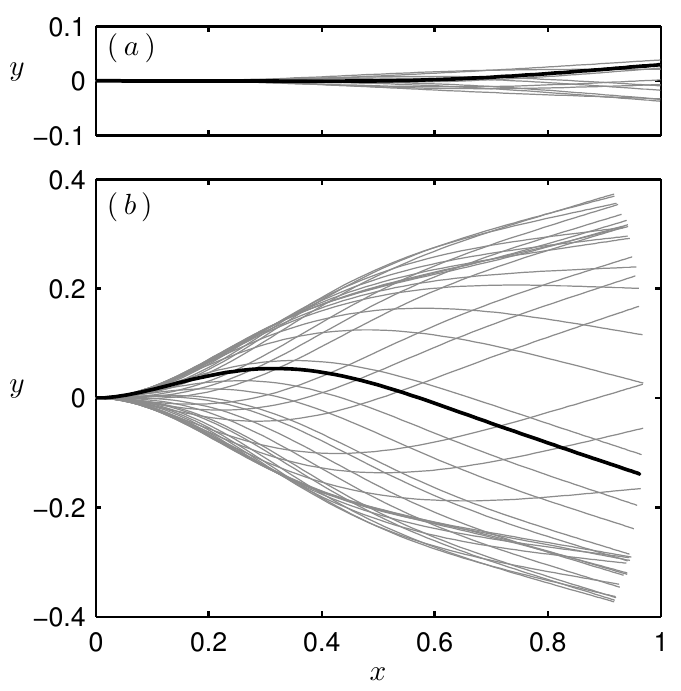}
\caption{\label{fig:flapping_flags}Flapping motion of flags at $\alpha=0.3$, $\beta=4$, $U^*=13$ and $(a)$ $\omega_0=3.25$ (no lock-in), $(b)$: $\omega_0=4.12$ (lock-in).}
\end{figure}

The reduced flow velocity is chosen at $U^*=13$\blue{, a value sufficiently higher than the critical velocity $U^*_c$.} The flag's behavior is observed to drastically differ with varying $\omega_0$ (Fig.~\ref{fig:flapping_flags}). When $\omega_0$ is within the range of destabilization, the flag undergoes a remarkably larger deformation (Fig.~\ref{fig:flapping_flags}$b$) than with other values of $\omega_0$ (Fig.~\ref{fig:flapping_flags}$a$).

\begin{figure}
\centering
\includegraphics{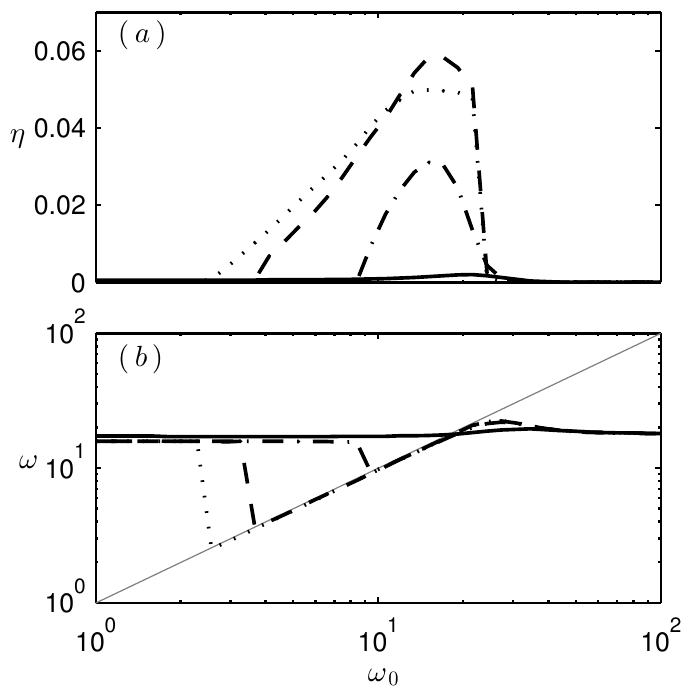}
\caption{\label{fig:eff_freq_omega0_4beta}$(a)$ Harvesting efficiency $\eta$ and $(b)$ flapping frequency $\omega$ as a function of $\omega_0$ for $\alpha=0.3$, $U^*=13$, and $\beta=0.05$ (solid), $\beta=1$ (dash-dot), $\beta=4$ (dashed), $\beta=8$ (dotted).}
\end{figure}

Figure~\ref{fig:eff_freq_omega0_4beta}$a$ shows the evolution of the efficiency with $\omega_0$, and demonstrates that this increased flapping amplitude indeed leads to a significant efficiency improvement: a maximum efficiency of $6\%$ is obtained here, significantly higher than the optimized efficiency obtained at $M^*=1$ and $U^*=13$ without inductance ($\sim 0.1\%$) \cite{michelin:2013}.

Within the high efficiency range, the flapping frequency is deviated and locks onto the natural frequency of the circuit, $\omega_0$ (Figure ~\ref{fig:eff_freq_omega0_4beta}$b$). A frequency lock-in is therefore observed here, similar to the classical lock-in observed in Vortex-Induced Vibrations (VIV) \cite{williamson:2004, grouthier:2013}: the frequency of an active oscillator (the flag) is dictated by the natural frequency of a coupled passive oscillator (the circuit). The lock-in range is extended by a reduction of the circuit's damping ($1/\beta$), consistently with what is observed in VIV for varying structural damping \cite{king:1973}. The lock-in range leading to high efficiency coincides with the range of $\omega_0$ 
associated 
with the destabilization by inductance (Fig.~\ref{fig:thrs_M1_4beta}). This suggests that a coupled piezoelectric flag, once destabilized by inductive effects, may flap at a frequency close to the natural frequency of the circuit. As a result, a permanent resonance takes place between the flag and the circuit, leading to increased flapping amplitude and harvesting efficiency.

\blue{
By varying $\omega_0$ within the lock-in range, Fig.~\ref{fig:eff_freq_omega0_4beta} shows that when the frequency of the output circuit matches the short-circuit natural frequency of the flapping flag ($\omega\sim 17.5$), the maximal efficiency is obtained for every value of $\beta$. This observation highlights again the interest of exciting piezoelectric structures at their natural frequencies for energy harvesting, as suggested by previous studies, where maximal efficiency is observed when the external forcing resonates with the piezoelectric system \cite{allen:2001, roundy:2005, ng:2005}. The existence of a lock-in extends this effect to a larger range of parameters, by maintaining the system at resonance, hence guaranteeing efficient energy transfers from the flag to the circuit.
}

\blue{
\section{Impact of piezoelectric coupling}
A decisive factor is the intensity of piezoelectric coupling, characterized by $\alpha$ in this work. It quantifies the proportion of the mechanical work transmitted to the circuit via piezoelectric effects. In practice, this coupling coefficient $\alpha$, defined in Eq.~\eqref{ParaAd}, depends primarily on the materials used for the piezoelectric flag. The importance of the coupling factor has been reported by many studies on energy harvesting by piezoelectric systems \cite{anton:2007, calio:2014piezoelectric}. In Refs.~\cite{doare:2011, michelin:2013}, a dependence of the harvesting efficiency to $\alpha^2$ was identified.

\begin{figure}
\centering
\includegraphics{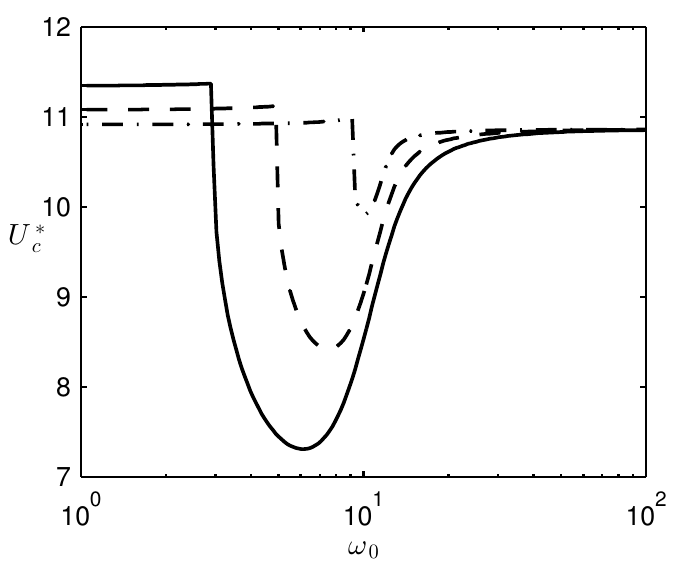}
\caption{\label{fig:threshold_3alpha_beta4}Evolution of the critical velocity with $\omega_0$ and $\beta=4$, $M^*=1$ and $\alpha=0.1$ (dash-dot), $\alpha=0.2$ (dashed) and $\alpha=0.3$ (solid).}
\end{figure}

\begin{figure}
\centering
\includegraphics{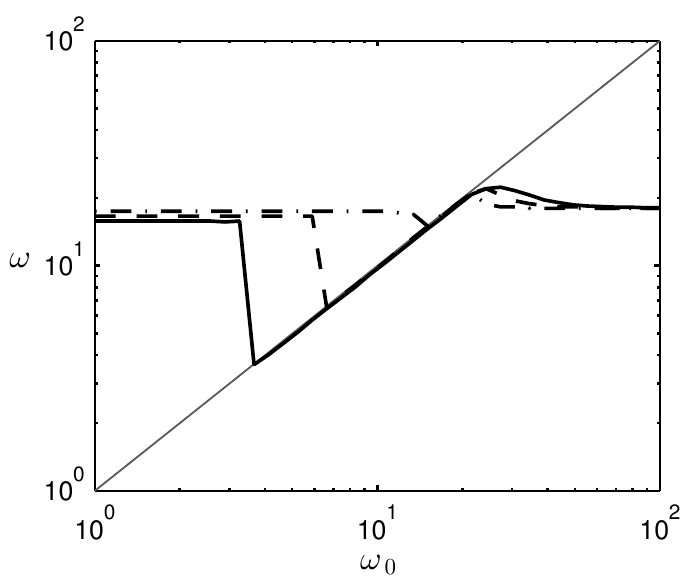}
\caption{\label{fig:freq_3alpha_beta4}Flapping frequency $\omega$ as a function of $\omega_0$ for $\beta=4$, $M^*=1$, $U^*=13$ and $\alpha=0.1$ (dash-dot), $\alpha=0.2$ (dashed) and $\alpha=0.3$ (solid).}
\end{figure}
In the present work, the influence of $\alpha$ is observed both in terms of critical velocity (Fig.~\ref{fig:threshold_3alpha_beta4}) and the lock-in range (Fig.~\ref{fig:freq_3alpha_beta4}). In Fig.~\ref{fig:threshold_3alpha_beta4}, lower critical velocities and larger destabilization range in terms of $\omega_0$ are observed with increasing $\alpha$. Fig.~\ref{fig:freq_3alpha_beta4} shows that the range of frequency lock-in also increases with increasing $\alpha$.
The impact of the coupling coefficient $\alpha$ on the system's performance is again identified: a strong piezoelectric coupling is desired so that the beneficial effects of a resonant circuit, namely the destabilization and the frequency lock-in, can be obtained.

In practice, $\alpha\sim 0.3$ could be expected with large scale devices \cite{doare:2011}, and in general $\alpha\sim 0.1$ is achievable, for example using two 10 cm $\times$ 10 cm MFC patches glued by an epoxy layer with a thickness of 0.1 mm. A potent piezoelectric material, leading to a strong coupling, is therefore an essential prerequisite to utilize the lock-in phenomenon.
}
\section{Conclusion}
The results presented here provide a critical and new insight on the dynamics of a piezoelectric energy-harvesting flag. First and foremost, they emphasize how the fundamental dynamics of the energy harvesting system and of the output circuit may strongly impact the motion of the structure and its energy harvesting performance. Also, they identify two major performance enhancements associated with the resonant behavior of the circuit, namely
\begin{inparaenum}[\upshape(i\upshape)]
 \item a destabilization of the fluid-solid-electric system, leading to spontaneous energy harvesting at lower velocity; and
 \item a lock-in of the fluid-solid dynamics on the circuit's fundamental frequency, resulting in an extended resonance and a significant increase of the harvested energy.
\end{inparaenum}

This lock-in behavior at the heart of both effects above is classically observed in VIV where it is also responsible for maximum energy harvesting \cite{grouthier:2014}; it is in fact a general consequence of the coupling of an unstable fluid-solid system to another oscillator's dynamics. We therefore expect that the conclusions presented in the present paper go beyond the simple inductive-resistive circuit considered here, and should be applicable to a much larger class of \blue{resonant systems. Such systems could be other forms of} electrical output circuits, \blue{or other} mechanical oscillators.

\blue{The critical impact of the coupling coefficient $\alpha$ on the system's performance is also underlined in the present work through its strong influence on the destabilization range and the lock-in range. The choice of the piezoelectric materials is therefore essential in the practical achievement of high efficiency.}

\begin{figure}
\centering
\includegraphics{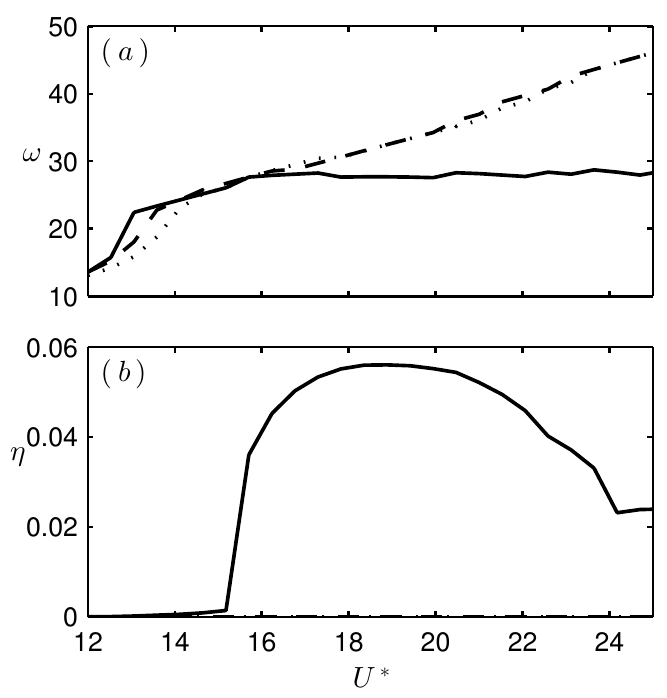}
\caption{\label{fig:freq_eff_beta3p4_3omega}$(a)$ Flapping frequency $\omega$ and $(b)$ harvesting efficiency $\eta$ as a function of $U^*$ for $\alpha=0.3$, $M^*=1$, $\beta=3.5$ and $\omega_0=0.1$ (dotted), $\omega_0=29$ (solid), $\omega_0=1000$ (dashed).}
\end{figure}

{This} lock-in mechanism also plays a critical role in the robustness of the energy harvesting process with respect to the flow velocity (Fig.~\ref{fig:freq_eff_beta3p4_3omega}). Lock-in indeed persists over a wide range of $U^*$, effectively acting as a passive control of the flapping frequency in response to flow velocity: while $\omega$ increases rapidly in the limit of weak fluid-solid-electric interactions, lock-in with the output circuit maintains $\omega\approx \omega_0$ and high harvesting efficiency over a large range of flow velocity. Such a control of the flapping frequency was shown to be essential for efficiency enhancement and robustness (e.g. to delay mode switches or frequency changes) \cite{michelin:2013}.

These results nonetheless illustrate the fundamental insights and technological opportunities offered by the full nonlinear coupling of 
\blue{a passive resonant system (electric, mechanical or other) to an unstable piezoelectric structure for the purpose of energy harvesting. The lock-in phenomenon and the enhanced performance demonstrated by the coupling between the piezoelectric flag and a simple resonant circuit open the perspective of applying different kinds of resonant systems to energy-harvesting piezoelectric system. The choice of an inductive circuit is motivated by its simplicity, while in practice, other designs of resonant circuits may present important advantages over the proposed formulation in our work.} 
Meanwhile, complex circuitry, such as propagative \cite{bisegna:2006} and active circuits \cite{lefeuvre:2006}, also represent interesting perspectives for future work.

\section*{Acknowledgment}
This work was supported by the French National Research Agency ANR (Grant ANR-2012-JS09-0017). S. M. also acknowledges the support of a Marie Curie International Reintegration Grant within the 7th European Community Framework Programme (PIRG08-GA-2010-276762).

%


%
%

%



\end{document}